\documentclass[12pt]{article}
\usepackage[pdftex]{hyperref,graphicx,color}

\newcommand\degree{^{\circ}}

\begin{document}

\includegraphics[width=0.2\textwidth]{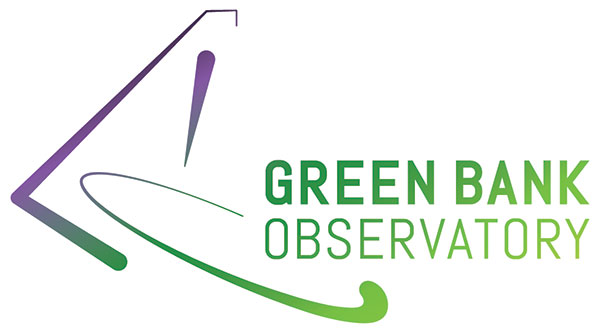}\hspace*{2cm}{\Large{\bf GBT
    Memo \#301}}

\begin{center}
\ \\
{\Large {\bf The GBT Gain Curve at High Frequency}}
\ \\
David T. Frayer, Frank Ghigo, \& Ronald J. Maddalena  (Green Bank Observatory)\\
2018 October 31
\end{center}

\begin{abstract}

  Recent measurements at Q-band (43 GHz) have verified the improved
  performance of the GBT provided by the updated gravity model that was
  deployed in the fall of 2014.  The measured gain curve is
  indistinguishable from 1.0 over an elevation range from 15$\degree$ to
  80$\degree$.  This represents a significant improvement on the previous
  gain curve from 2009 that showed decreasing efficiencies below
  40$\degree$ and above 65$\degree$ elevation.   The current estimated
  surface errors, under good conditions, is 230\,$\mu$m for the GBT.

\end{abstract}



\section{Background}

The performance of the Robert C. Byrd Green Bank Telescope
(GBT)\footnote{The Green Bank Observatory is a facility of the
  National Science Foundation under cooperative agreement by
  Associated Universities, Inc.} has improved significantly over the
years with updates to the active surface model.  Using only the GBT
finite element model (FEM) for the surface, the aperture efficiency
falls off from the peak efficiency value at an elevation of about
$52\degree$ due to large-scale gravitation deformations (Condon 2003;
Balser, Prestage, \& Nikolic 2005) [see dotted-line in Fig. 1].  Using
the FEM model in combination with a Zernike-gravity model derived from
out-of-focus holography (OOF) measurements of astronomical sources
(Nikolic, Balser, \& Prestage 2006; Nikolic et al. 2007) has vastly
improved the performance of the GBT.  After making the appropriate
mechanical corrections to many of the 2209 surface actuators based on
data from the 12 GHz holography system installed on the GBT (Schwab
2008), the effective surface error was improved from about 390\,$\mu$m
to 240\,$\mu$m (Hunter et al. 2011).  Before 2009, the GBT surface
was not accurate enough for efficient observations at 3mm wavelengths,
but after the implementation of Zernike-gravity model and using OOF
measurements to derive residual ``Thermal'' Zernike coefficients for
the current conditions of the telescope, aperture efficiencies of
about 40\% are obtainable.

After the improvements in 2009, the measured Q-band gain curve of the
GBT was still not optimal (2009b dashed line in Figure~1).  The 2009
gain curve is based on the 2005WinterV2 Gravity model.  This model was
replaced by the 2010WinterV1 model, but an updated Q-band gain curve
was not derived for this model, and the results from the 2009 gain
curve have continued to be used.  After finishing the replacement of
the sub-reflector actuators in 2013, a updated Zernike-gravity model
(2014FallV1) was derived based on all AutoOOF observations after the
actuator replacements (from Nov. 2013 through the the fall of 2014,
Maddalena et al. 2014).  The advantage of this model is that it was
derived from many sets of observations which effectively averaged over
a range of temperatures and conditions for the GBT.  The 2014FallV1
model has proven successful during the 2015 through 2018
high-frequency observing seasons.  The typical residual AutoOOF
surface rms corrections have been small (100--250$\mu$m) indicating
the appropriateness of the model.  In this memo, we derive a new
Q-band gain curve for the GBT based on measurements using the updated
2014 gravity model.

\begin{figure}[tbh]
\begin{center}
\includegraphics[width=0.8\textwidth]{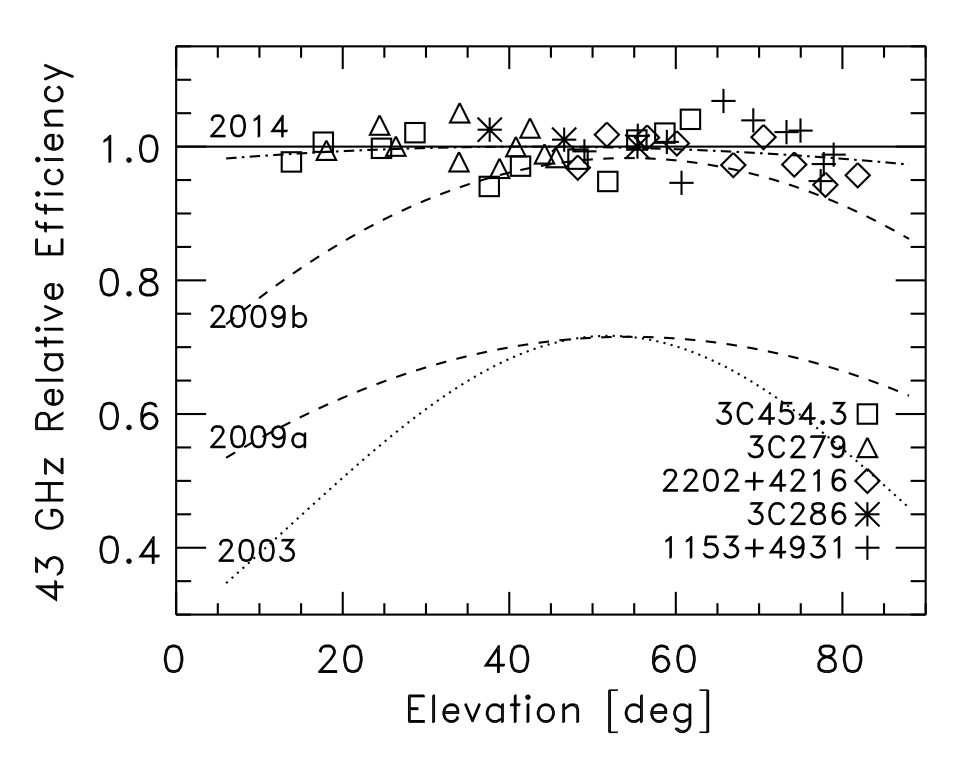}
\end{center}
\vspace*{-5mm}

\caption{The relative aperture efficiency of the GBT as a function of
  elevation at 43 GHz.  The data points are for observations carried
  out with the updated 2014 Zernike-gravity model.  The dashed-dotted
  line shows a 2nd order polynomial fit to the data and highlights the
  improvement provided by the 2014 Zernike-gravity model in comparison
  to the previous gain curve (2009b).  The dashed lines show the
  improvement of the surface carried out in 2009.  The 2009a dashed
  line represents a surface rms of error of 390\,$\mu$m, while the
  2009b dashed line has an rms error of 240\,$\mu$m.  The 2003 dotted
  line shows the performance without the Zernike-gravity model for
  comparison.}

\end{figure}

\section{Results}

\subsection{Q-band Gain Curve}

Figure~1 shows the results of the Q-band (43 GHz) gain-curve
observations as a function of elevation.  The data for 3C454.3
(2253+1608), 3C286, and 2202+4216 were taken as part of the Q-band
program TGBT18A\_503\_01 observed 2018 May 25.  All data have been
corrected for the atmosphere and placed on the $T_{A}^{\prime}$
temperature scale.

\begin{equation}
T_{A}^{\prime} = T_{A} \exp(\tau_{o}/\sin({\rm El})),
\end{equation}
where $T_{A}$ is the observed antenna temperature of the source,
$\tau_{o}$ is the zenith opacity derived from the local weather
database, and El is the elevation of the source.  To avoid
uncertainties associated with absolute flux calibration, the
measurements for each source were normalized to the average value
observed within the intermediate elevation range of
45$\degree$--60$\degree$ where the gain of the telescope is expected
to be optimal.  We found no significant drop in telescope efficiency
at low elevation as previously seen in the 2009 gain curve.  The
results are consistent with a flat gain curve from 15$\degree$ to
80$\degree$ elevation.  We do not have sufficient data to derive the
gain curve above 80$\degree$ or below 15$\degree$.  To confirm these
results, we collected archival Q-band observations of 3C279 and
1153+4931 which are also plotted in Figure~1.

To facilitate comparisons with previous results, the data were fitted
with a 2nd order polynomial as a function of zenith angle.

\begin{equation}
{\rm Gain(ZD)} = {\rm A0} + {\rm A1}{\rm(ZD)} + {\rm A2}{\rm(ZD)}^{2},
\end{equation}
where ZD is the zenith angle in degrees, and A0, A1, and A2 are the
fitted polynomial coefficients.  Table~1 gives the fitted parameters
along with their errors for the 2014 Zernike-gravity model.  The gain
curve has been normalized to a maximum value of 1.0.  For comparison
the previous 2009 coefficients are also tabulated.  The gain-curves
based on these coefficients are shown in Figure~1.

\begin{table}[tbh]
\caption{Gain Curve Coefficients for 43 GHz}
\centering
\begin{tabular}{lccc}
\hline
\hline
      &A0 & A1& A2\\
\hline
2009 & 0.8618 & $7.737\times10^{-3}$ & $-1.0838\times10^{-4}$\\ 
2014  &0.971$\pm$0.020& $(1.24\pm1.06)\times
10^{-3}$&($-1.31\pm1.28)\times10^{-5}$\\
\hline
\end{tabular}
\end{table}

The average for all data plotted in Figure~1 is 0.997 with
standard-deviation scatter of 0.030 (which implies a $1\sigma$ 3\%
uncertainty for an individual data point).  The observational
uncertainties derived here are less than those found for previous GBT
Q-band gain-curve analyses (e.g., see gain-curve plot presented in
Hunter et al. 2011, and the GBT PTCS wiki pages which typical report
15\% observational errors).  Each point plotted in Figure~1 represents
an average of the two polarizations and for two peak scans.  The
normalization of values per source and per observing session also
reduced the observational scatter significantly.  The derived gain
curve is independent of the absolute calibration of the source and
aperture efficiency of the telescope at the time of observation.

\begin{figure}[tbh]
\begin{center}
\includegraphics[width=0.8\textwidth]{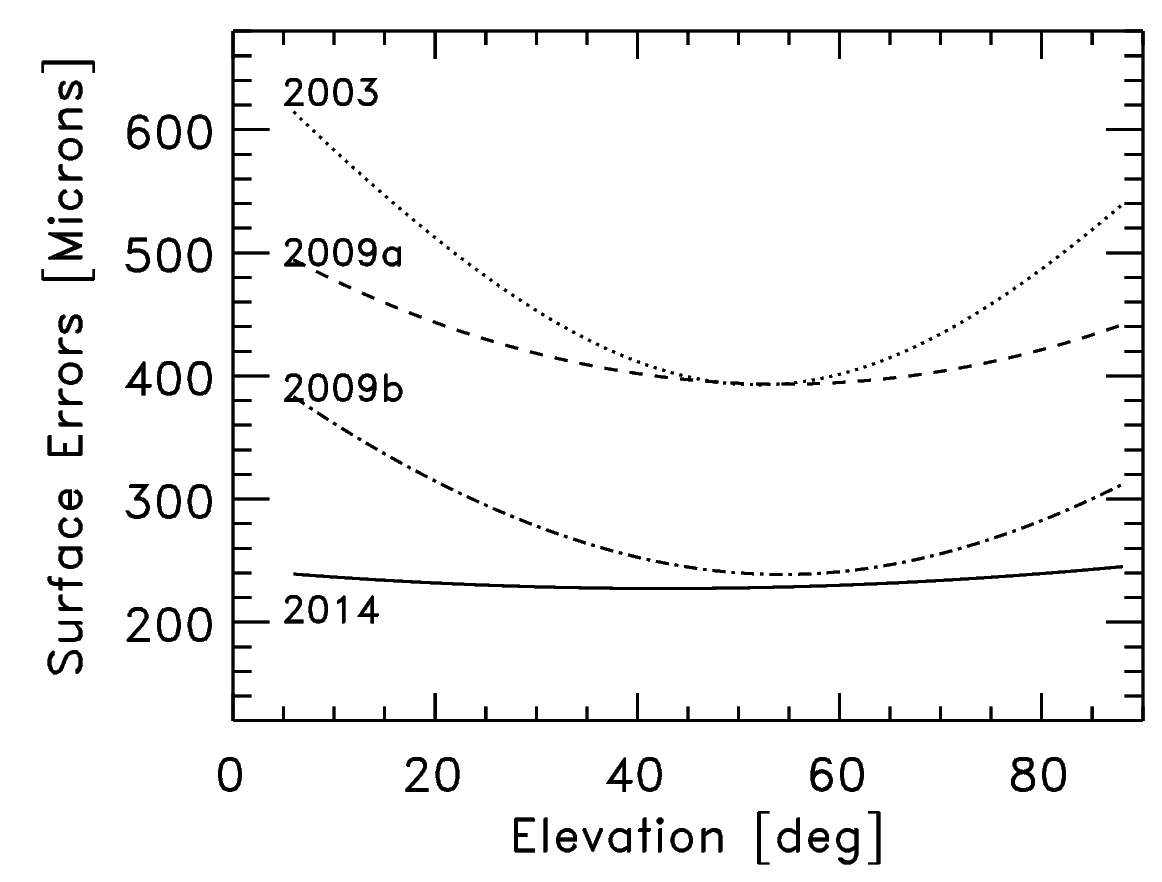}
\end{center}
\vspace*{-5mm}

\caption{The effective surface errors for different GBT surface models
  over the years based on observations as a function of elevation at
  43 GHz and assuming the Ruze equation (Equation~4). }

\end{figure}

\begin{table}[tbh]
\caption{GBT Surface Errors and Peak Aperture Efficiency}
\centering
\begin{tabular}{lcccccc}
\hline
\hline
     &Surface Error&\multicolumn{5}{c}{Aperture Efficiency}\\
     &[$\mu$m]& 10\,GHz &30\,GHz&43\,GHz&80\,GHz&110\,GHz\\ 
\hline
2003 &390& 69\%&56\%&43\%&13\%&3\%\\
2009a&390& 69\%&56\%&43\%&13\%&3\%\\
2009b&240& 70\%&65\%&59\%&37\%&21\%\\
2014 &230& 70\%&65\%&60\%&39\%&23\%\\
\hline
\end{tabular}
\end{table}

\subsection{GBT Aperture Efficiency}

The derived gain curve at 43 GHz can be used to predict the
performance of the telescope at other frequencies as a function of
elevation.  Based on observations of sources with known flux density
($S_{\nu}$), the derived aperture efficiency ($\eta_{a}$) for the GBT
is
\begin{equation}
\eta_{a} = 0.352 T_{A}^{\prime}/S_{\nu}.
\end{equation}
The aperture efficiency is related to the surface errors using the
Ruze equation: 
\begin{equation}
\eta_{a} =0.71 \exp[-(4\pi\epsilon/\lambda)^2)],
\end{equation}
where the coefficient 0.71 is the aperture efficiency at long
wavelengths for the GBT and $\epsilon$ is the rms uncertainty of the
surface.  Based on 43 GHz observations, Table~2 shows the derived
aperture efficiency and corresponding surface errors for the telescope
over time for the GBT.  The Ruze equation was used to scale the 43 GHz
results to other frequencies (Table~2).

The efficiencies given in Table~2 are for the optimal elevation.  The
gain curves shown in Figure~1 have been used to derive the effective
surface errors as a function of elevation (Figure~2), which are then
used to derive the aperture efficiency as a function of elevation and
frequency using the Ruze equation.  These results are plotted in
Figures~3-7.

\section{Discussion}

Previously, there were concerns that the AutoOOF solutions may not be
applicable when observing sources at different elevations due to
inadequacies of the Zernike-gravity model (see comments within the
PTCS wiki pages over the last decade).  The AutoOOF associated with
the recent Q-band gain-curve observations (TGBT18A\_503\_01) was done
at high elevation (80 deg), and we found no decrease in efficiency at
low elevation.  These results imply that the surface corrections from
an AutoOOF at one elevation are applicable at other elevations, when
using a good Zernike-gravity model.

The 2014FallV1 Zernike-gravity model used for this memo is vastly
superior to the 2005WinterV2 Zernike-gravity model that was used for
the 2009 gain-curve derivation.  Unfortunately, the performance
provided by the 2010WinterV1 Zernike-gravity model was never
quantified with an accurate Q-band gain curve.  However, based on
archival data its performance is much closer to the the 2014FallV1
model than the 2005WinterV2 model.  The deprecated 2014WinterV1 model
should be avoided, since this model had poor performance.

\section{Concluding Remarks}

The observed 43 GHz gain curve is flat a function of elevation which
validates the 2014FallV1 Zernike-gravity model.  The performance of
the GBT over the years has continued to be improved as refinements
have been made to the active surface model.  These improvements have
made observations more efficient and have enabled 3mm observations
with the GBT, which has motivated the development of new
instrumentation on the GBT that operates within this band (e.g.,
Argus, Mustang-2, and the 4mm Receiver).

The observatory has recently received funding for an optical laser
scanner that would permit more rapid monitoring of the surface (Green
Bank Observatory News 2018).  Once implemented this should help to
maintained an accurate surface during long sets of observations and
potentially permit efficient day-time observing for 3mm observations.

\section{References}

\begin{enumerate}

\item Condon, J. 2003, GBT Efficiency at 43 GHz, GBT PTCS Project
  Note\#31

\item Balser, D., Prestage, R. M., \& Nikolic, B. 2005, GBT Aperture
  Efficiency at Q-band (43 GHz), GBT PTCS Project Note\#43
 

\item Hunter, T. R., Schwab, F. R., White, S. D., et al. 2011, Holographic
  Measurement and Improvement of the Green Bank Telescope Surface,
  PASP, 123, 108

\item Maddalena, R. J., Frayer, D. T., Lashley-Colhirst, N. \& Norris,
  T. 2014, The Updated 2014 Gravity Model, GBT PTCS Project Note\#76

\item Nikolic, B., Balser, D. S., \& Prestage, R. M. 2006,
  Out-Of-Focus Holography at the Green Bank Telescope, The
  Gain-Elevation Curve and Absolute Efficiency, GBT PTCS Project
  Note\#47

\item Nikolic, B., Prestage, R. M., Balser, D. S., Chandler, C. J., \&
  Hills, R. E 2007, Out-of-Focus Holography at the Green Bank
  Telescope, A\&A, 465, 685

\item Schwab, F. R. 2008, Summer/Autumn 2006 GBT Panel Corner Setting
  Measurements, GBT PTCS Project Note\#61

\end{enumerate}


\begin{figure}[tbh]
\begin{center}
\includegraphics[width=0.9\textwidth]{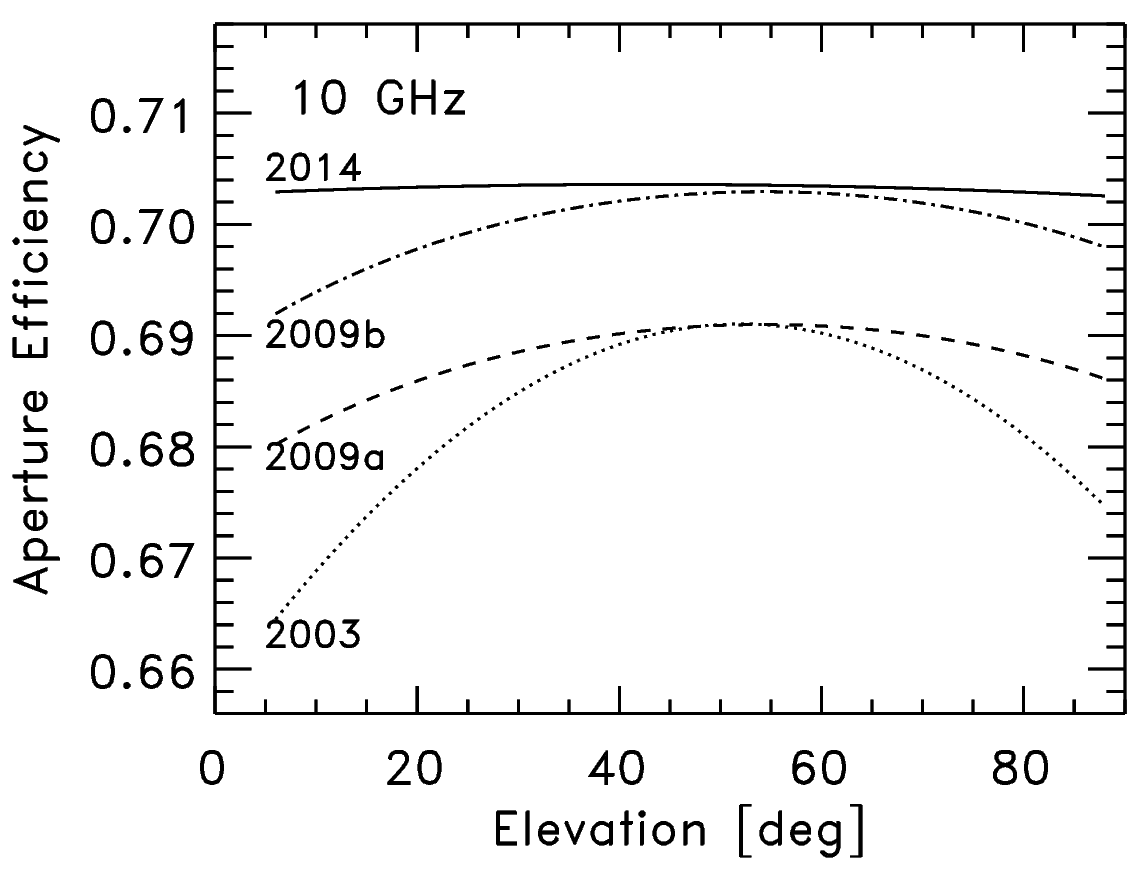}
\end{center}
\vspace*{-5mm}

\caption{The effective aperture efficiency of the GBT at 10 GHz as a
  function of elevation for different GBT surface models over the
  years.  The curves are based on the surface errors computed for
  Figure~2 and assume the Ruze equation (Equation~4). }

\end{figure}

\begin{figure}[tbh]
\begin{center}
\includegraphics[width=0.9\textwidth]{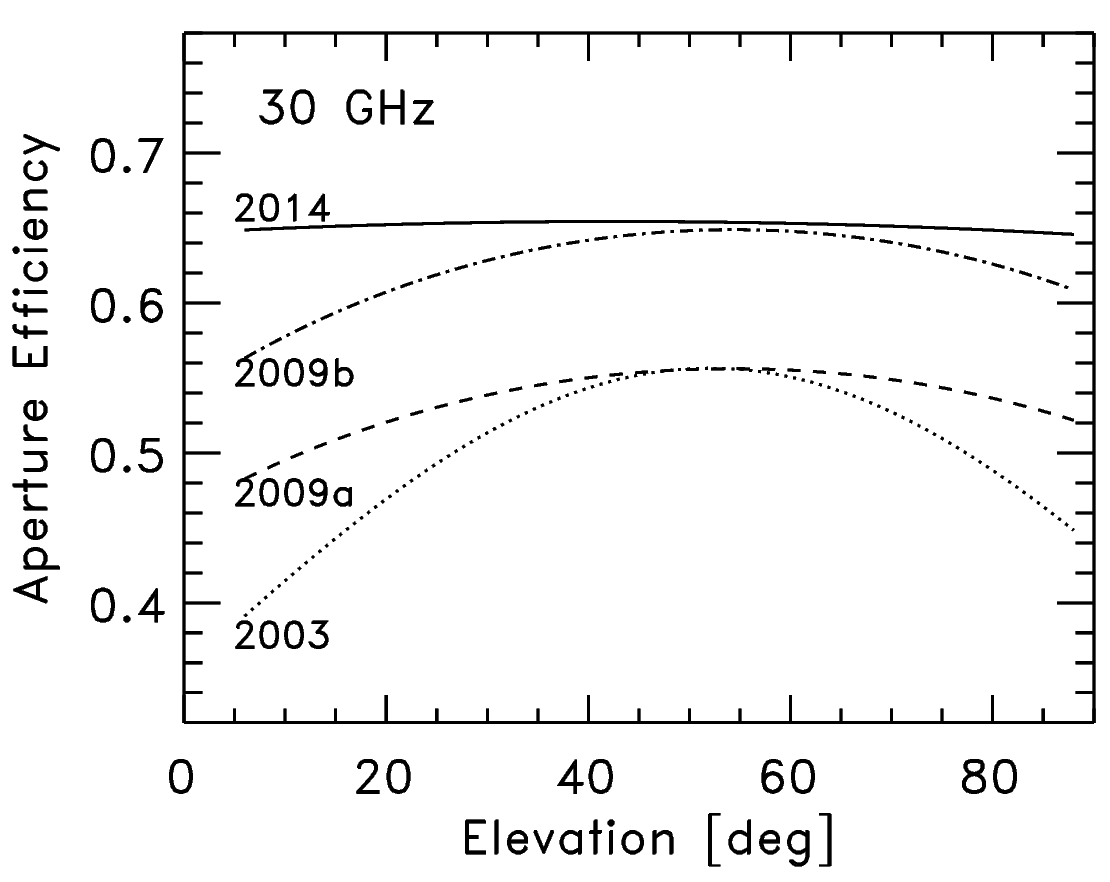}
\end{center}
\vspace*{-5mm}

\caption{The effective aperture efficiency of the GBT at 30 GHz as a
  function of elevation for different GBT surface models over the
  years.  The curves are based on the surface errors computed for
  Figure~2 and assume the Ruze equation (Equation~4). }

\end{figure}

\begin{figure}[tbh]
\begin{center}
\includegraphics[width=0.9\textwidth]{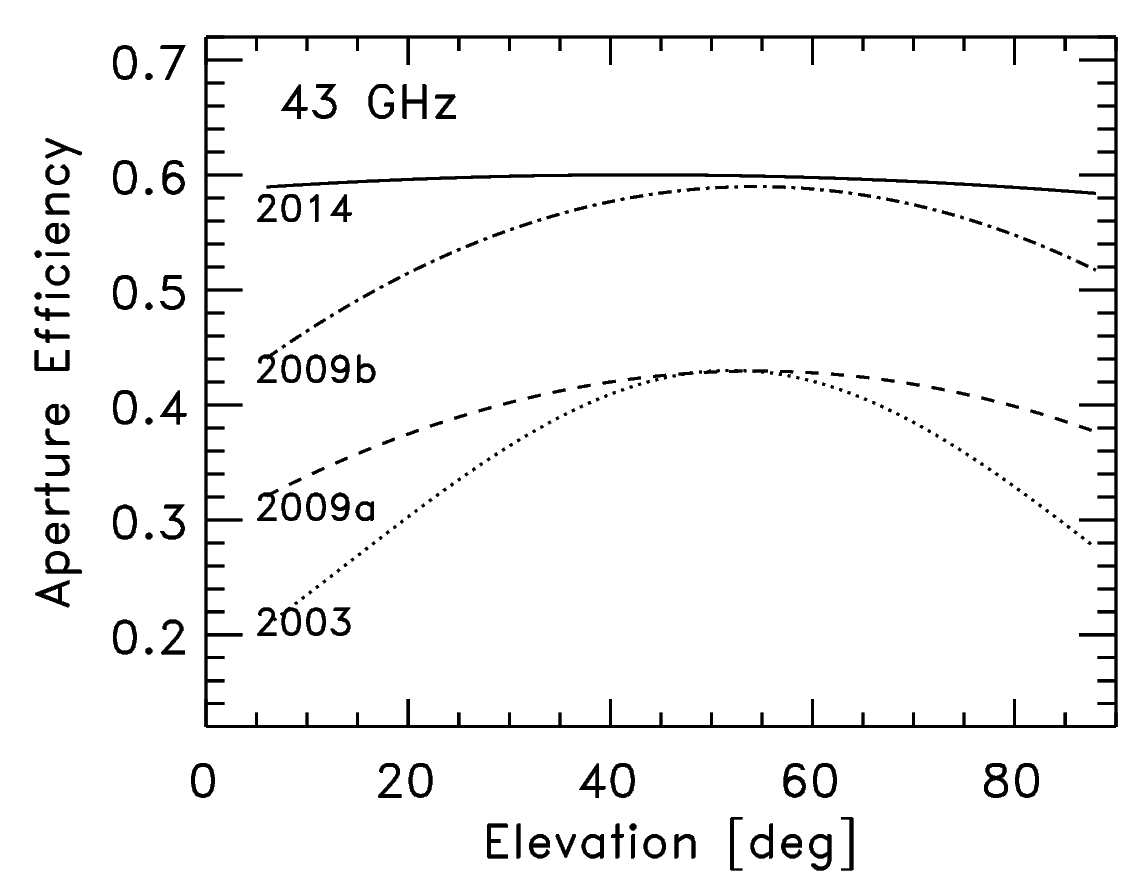}
\end{center}
\vspace*{-5mm}

\caption{The effective aperture efficiency of the GBT at 43 GHz as a
  function of elevation for different GBT surface models over the
  years.  The curves are based on the surface errors computed for
  Figure~2 and assume the Ruze equation (Equation~4). }

\end{figure}

\begin{figure}[tbh]
\begin{center}
\includegraphics[width=0.9\textwidth]{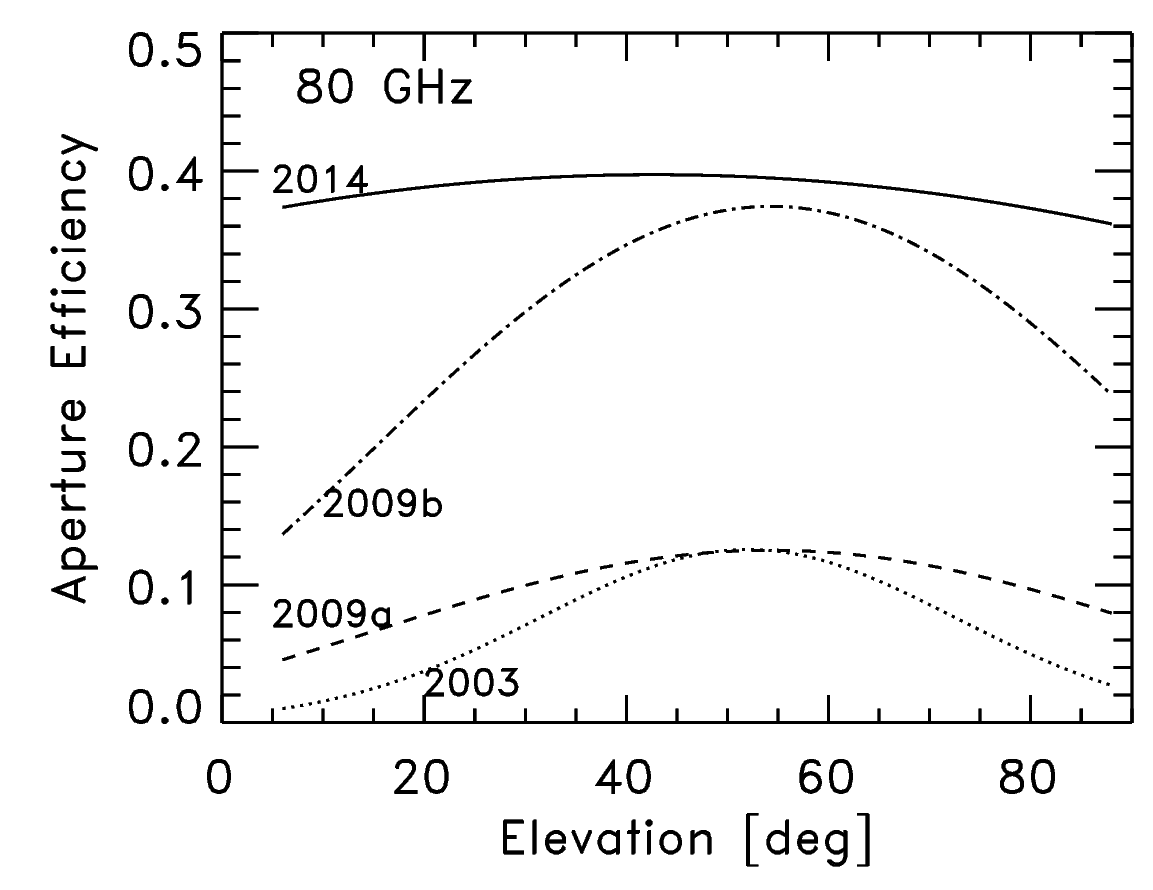}
\end{center}
\vspace*{-5mm}

\caption{The effective aperture efficiency of the GBT at 80 GHz as a
  function of elevation for different GBT surface models over the
  years.  The curves are based on the surface errors computed for
  Figure~2 and assume the Ruze equation (Equation~4). }

\end{figure}

\begin{figure}[tbh]
\begin{center}
\includegraphics[width=0.9\textwidth]{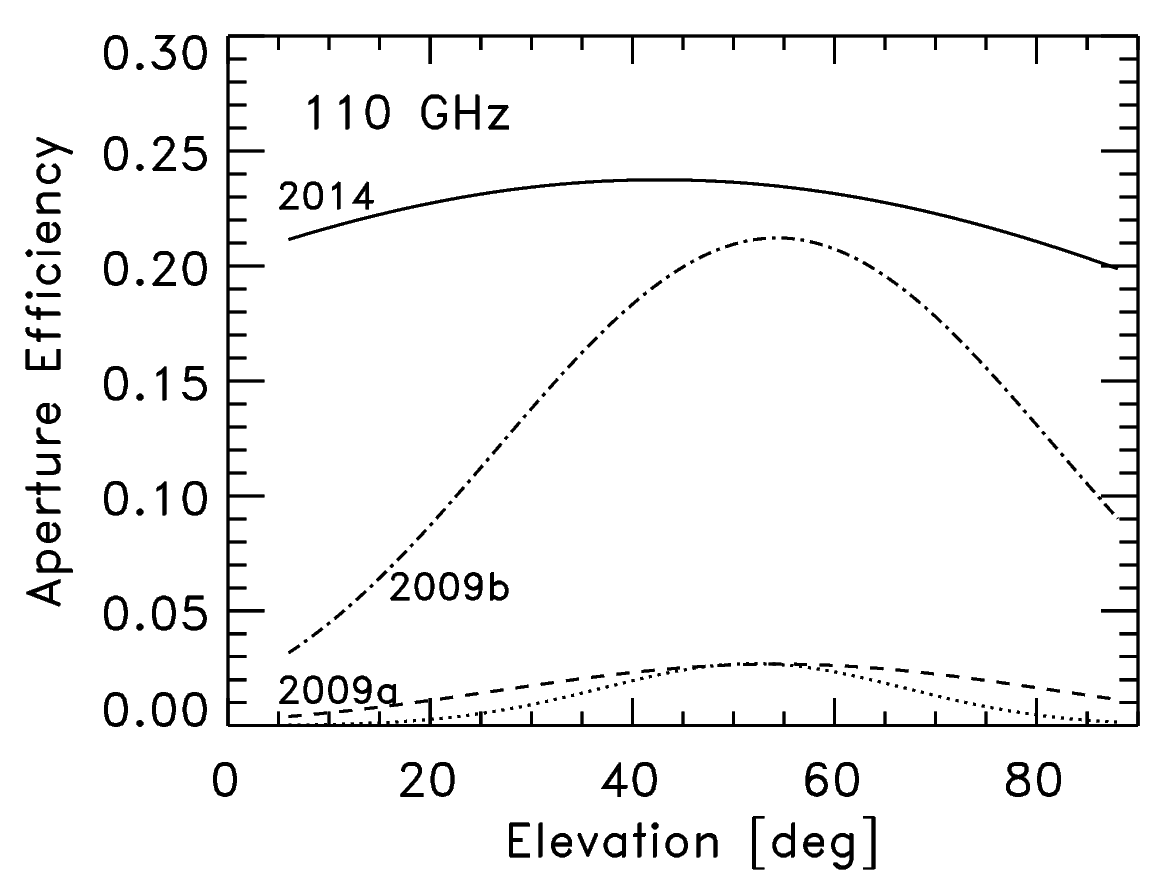}
\end{center}
\vspace*{-5mm}

\caption{The effective aperture efficiency of the GBT at 110 GHz as a
  function of elevation for different GBT surface models over the
  years.  The curves are based on the surface errors computed for
  Figure~2 and assume the Ruze equation (Equation~4). }

\end{figure}

\end{document}